\documentclass[fleqn,twoside]{article}
\usepackage[headings]{espcrc2}

\readRCS
$Id: espcrc2.tex,v 1.2 2004/02/24 11:22:11 spepping Exp $
\usepackage{epsfig}

\newcommand{\AmS}{{\protect\the\textfont2
  A\kern-.1667em\lower.5ex\hbox{M}\kern-.125emS}}

\title{Carpet-3 - a new experiment to study primary composition around the knee}

\author{J. Szabelski for the Carpet-3 collaboration}

\runtitle{Carpet-3 - a new experiment to study primary composition around the knee} \runauthor{J.
Szabelski}

\begin{document}

\begin{abstract}
We propose a new experiment to study primary composition around the knee. The Carpet-3 EAS array is the further
development of the Carpet-2 EAS array (1700 m a.s.l., Baksan Valley) and it is supposed to be a
multi-component and multi-purpose array detecting, in the EAS's with $E > 10^{13}$ eV,   electrons, gammas,
muons (with a threshold energy of 1 GeV), hadrons (with energies more than 30 GeV), and thermal neutrons as
well. The experimental data are to be used in the multi-component analysis to make conclusions about the
composition of the primary cosmic rays. \vspace{1pc}
\end{abstract}

\maketitle

\section{Introduction}
The nature of the knee in the primary cosmic ray spectrum at $E_k \approx 3\times 10^{15}$ eV, where
the power law index $\gamma$ is changed from $\gamma \approx 2.7$ to $\gamma \approx 3.1$, is one of
main problems in modern high energy astrophysics. Different models for acceleration and propagation of
high-energy cosmic rays in the Galaxy predict different shapes of the knee. Especially intriguing is
the fact  that the knee energy is very close to the maximum attainable energy for protons accelerated
in the shock front of SNRs \cite{Ptuskin2004}. Today's view is that the origin of the knee could
provide an answer to the puzzling problem of cosmic ray origin.

The knee in the shower size spectrum at about $10^6$ particles was first observed by the MSU group in
1958 \cite{Kulikov1958} and at the same time the astrophysical interpretation of the knee, namely, the
steepening of the primary cosmic ray spectrum, was proposed. Since then many experiments   have been
carried out, and breaks in the size spectra of Extensive Air Shower (EAS) have been found for its
different components: electromagnetic (e.m.), muonic, Cherenkov light, and hadronic. It is obvious
inconsistency  between characteristics of these breaks that does  not allow us to give a final
conclusion concerning the shape of the knee in the primary spectrum \cite{Shatz2003}. What is more,
the dependence of the mean atomic number of primary cosmic rays on primary energy around the knee
obtained in different experiments have an enormous scattering - from pure protons to pure irons. These
contradictions gave rise to different alternative explanations of the knee: e.g., the knee could be a
consequence of some change in the hadronic interaction properties \cite{Petrukhin2001} or just a
methodical effect \cite{Stenkin 2003}.

Last decade the study of the knee was carried out with modern EAS arrays able to simultaneously
register different EAS components and the EAS inverse approach for reconstruction of primary energy
spectra has been developed (see \cite{Garyaka2007} and references therein). It should be noted that
this approach is based on different  model--dependent EAS simulations with the effect that the
reconstructed primary energy spectra differ distinctly. It is quite disappointing to see that the
primary energy spectra reconstructed at different EAS arrays differ even when obtained with the same
interaction model. Usually modern EAS arrays measure characteristics for two EAS components: the muon
(GRAPES-3, GAMMA) or hadronic (Tibet) components are registered in addition to the electromagnetic
one.  With the exception of the KASCADE EAS array which can register both muon and hadron components
in addition to the electromagnetic one, however for reconstruction of primary mass group energy
spectra only data from electromagnetic and muon component were used~\cite{FzkMassMerida}. Thus, it is
obvious that solution of the knee problem demands an array measuring all possible EAS parameters with
high precision.

\section{Carpet-3 EAS array}
The Carpet-3 EAS array (Fig.1) is the further development of the Carpet-2 EAS array
\cite{Dzhappuev2007}. The central part of the array (the Carpet proper) consists of 400 individual
liquid scintillation detectors of 0.5 $m^2$ each. Six outside points have 18 scintillation detectors
each. The signals from the latter are used for timing and for EAS's arrival direction reconstruction.

\vspace{5pt} \centerline { \epsfig{file=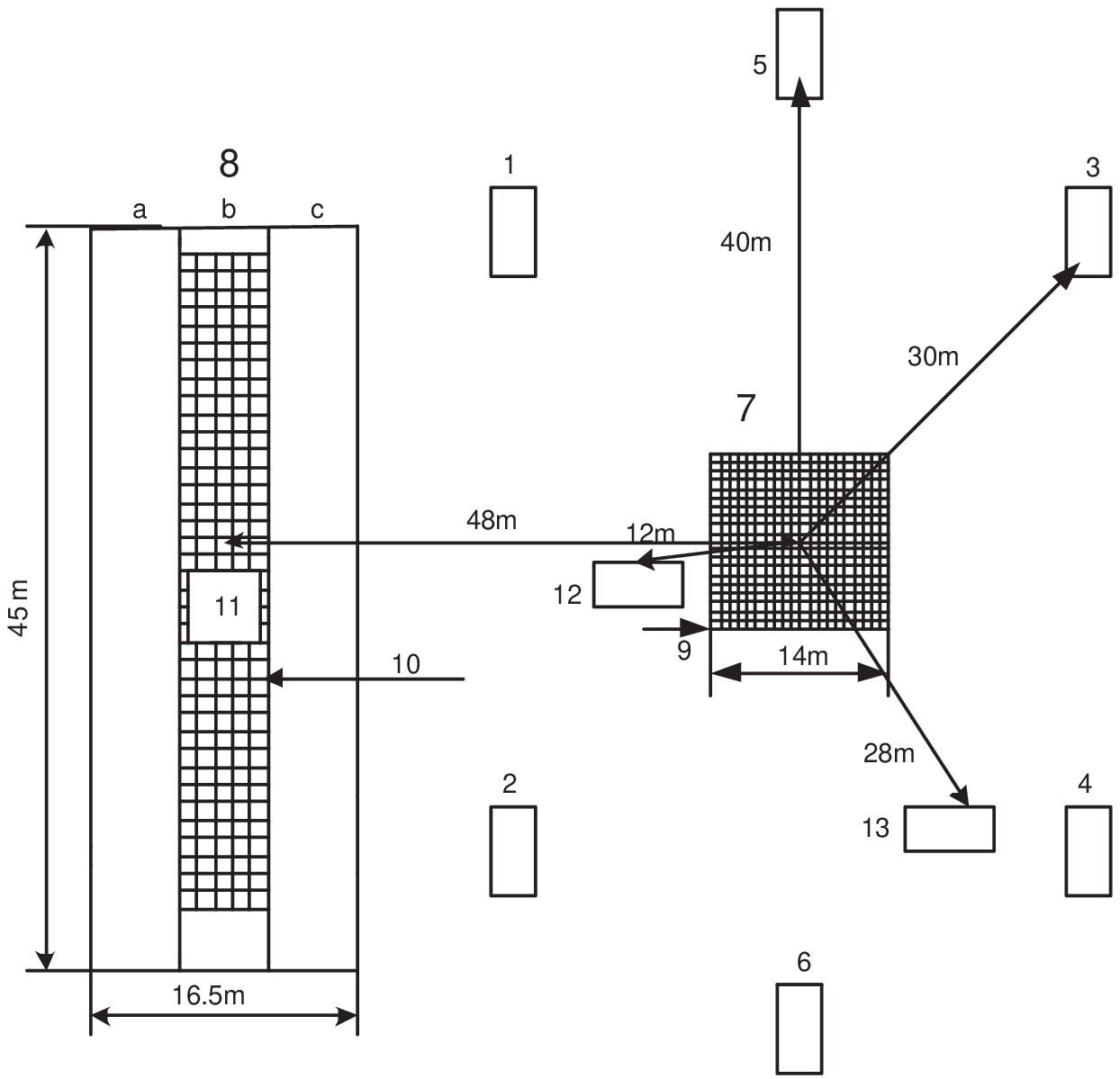,width=7cm}}
 {\small \sl Fig.1. Carpet-3 EAS array.
1-6  -  outside points, 7  -  Carpet,  8 - planned Muon  Detector (MD) (10 - present MD),
 11-13 - outside points with thermal neutron detectors (TND).}
 \vspace{5pt}

Muon detector (MD) is situated in the underground tunnels under 2.5 m of soil absorber (500 $g/cm^2$).
The distance between MD and Carpet centers is 48 m. At present MD consists of 175 plastic
scintillation detectors placed in the central tunnel. Each detector has an area of 1 $m^2$. The
detectors are attached to  the ceiling of the underground tunnel. MD's area is supposed to be enlarged
from $175 m^2$ to $630 m^2$, the effect of such enlargement is presented in Fig.2.

The important part of the array will be the thermal neutron detectors (TND) \cite{Stenkin2008}. We use
a thin layer of a mixture of old inorganic scintillator ZnS(Ag) with LiF enriched in $^6$Li up to
90\%. Thermal neutrons are recorded due to $^6Li(n,\alpha)^3H + 4.78$ MeV reaction. ZnS scintillator
is the best scintillator for heavy particle detection and it produces $\sim 160000$ light photons per
one captured neutron.

\vspace{5pt} \centerline { \epsfig{file=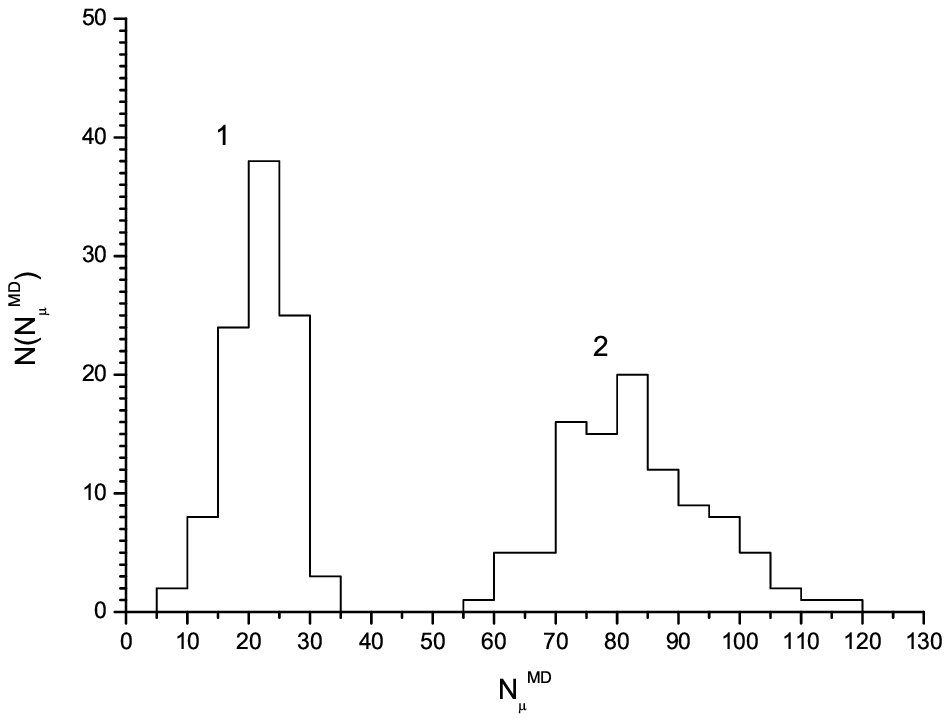,width=8cm}}
 {\small \sl Fig.2. Distribution of the number of muons in MD  for the present ($S=175 m^2$)
  (labelled \lq 1') and planned ($S=630 m^2$) (labelled \lq 2') MD's area for
  primary iron nuclei of $E_0 = 10^{15}$ eV with EAS
  axes – in Carpet center (CORSIKA, QGSJet01 + GHEISHA, \cite{Heck98}).
  For the present MD's area: $\overline N_{\mu}^{MD} = 21.3$, $\sigma (N_{\mu}^{MD}) = 4.8$,
   $\frac{\sigma (N_{\mu}^{MD})}{\overline N_{\mu}^{MD}} = 0.23$.
For the planned MD's area: $\overline N_{\mu}^{MD} = 82.4$, $\sigma (N_{\mu}^{MD}) = 11.9$,
   $\frac{\sigma (N_{\mu}^{MD})}{\overline N_{\mu}^{MD}} = 0.14$}
 \vspace{5pt}

 It means one could make a large detector viewed by a single PMT and have enough
light to register neutrons. In our case we have $\sim 50$ photo-electrons from PMT photo-cathode. The
efficiency of thermal neutron detection was found to be 20\%. Pulse duration (the fastest component)
is equal to $\sim 40$ ns. Taking into account that heavy particles also excite slower component one
can use this for pulse shape selection. The scintillator layer is very thin  (30 $mg/cm^2$) so it is
almost insensitive to single charged particles and  gamma--rays, but it can be successfully used for
EAS particle density measurements. The array's response simulation will be performed in order to
optimize the number of TND's and their arrangement.

\section{Carpet-3 performance capabilities}
The EAS's with axes well inside the Carpet will be analyzed. Due to large continuous area of the
Carpet and relatively small areas of individual detectors the accuracy of axis position for such
events is $\sim 0.2$ m. This gives a possibility to perform fine-resolution measurements of the
lateral distribution function (LDF) and its fluctuations near the EAS core. For example, the measured
LDF of the charged particles for $N_e \ge 2\cdot 10^5$ ($\overline N_e = 5\cdot 10^5$) and simulated
ones are compared in Fig. 3.

\vspace{5pt} \centerline { \epsfig{file=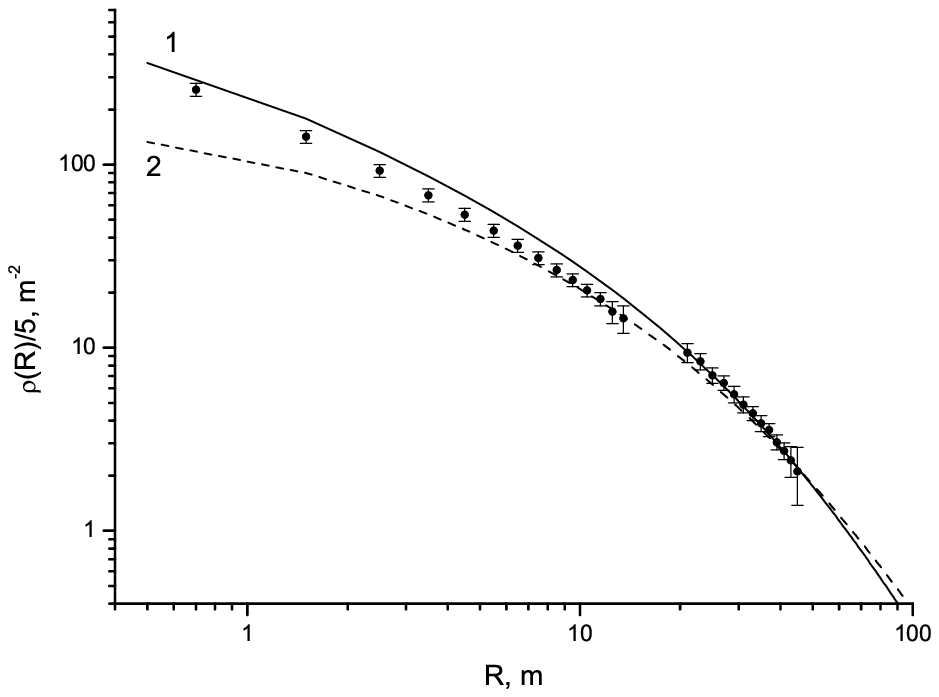,width=8cm}}
 {\small \sl Fig.3. LDF of the charged particles
  in the Carpet experiment (points). Lines - simulation (CORSIKA, QGSJet01c + GHEISHA, \cite{Heck98}):
   1 - primary protons,
   2 - primary iron nuclei.}
 \vspace{5pt}

The number of thermal neutrons, as a first approximation, is proportional to the total  number of
hadrons in EAS. The total  number of hadrons, in turn, can be used as an energy estimator of a primary
particle (Fig.4), and to this effect the total area and location of thermal neutron detectors (TND)
will be optimized.

In addition to muons with $E_{\mu} > 1$ GeV, MD can also detect hadrons with $E_h \ge 30$ GeV. In the
absorber above MD the EAS hadrons generate cascades producing energy deposit in the scintillator. The
thickness of the absorber is equal to $\sim 20$ radiation lengths; it is enough to absorb
electromagnetic component but not the hadronic cascades because of its having only $\sim 5$ hadron
interaction lengths. The number of such hadrons can be taken as the number of cascades in MD, i.e.,
the number of spots with local density of $\ge 10$ particles per $m^2$
\cite{Dzhappuev2007b,Dzhappuev2008}. It is not only the number of hadrons in MD but their total energy
as well that can be measured. For this purpose the TND's will be placed under the scintillator
detectors, in the underground tunnels, to measure the number of thermal neutrons. Because the number
of thermal neutrons depends on energy of cascades, the total energy of hadrons in MD can be measured
(the method of ionization-neutron calorimeter, developed for the INKA project \cite{Aleksandrov2003}).

\vspace{5pt} \centerline {\epsfig{file=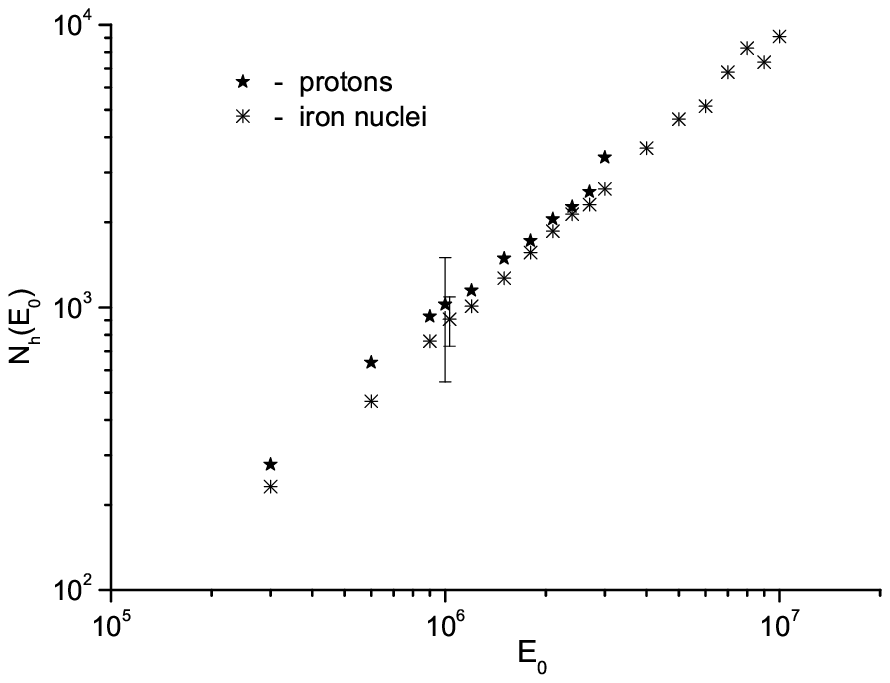,width=8cm}} {\small \sl Fig.4. Dependence of mean
number of hadrons with $E_h \ge 30$ MeV on energy of primary
 protons and iron nuclei (CORSIKA, QGSJet01c + GHEISHA, \cite{Heck98}).
}
 \vspace{5pt}

\section{CONCLUSIONS}
For each EAS will be measured, at least, 6 parameters:

1)$N_{ch}^C$  - the number of charged particles in "Carpet";

2)$s^C$   - age parameter of NKG-function near the EAS axis;

3)$N_{n}^{ND}$  - the number of thermal neutrons in the surface TND's;

4)$N_{\mu}^{MD}$  - the number of muons ($E_{\mu} > 1$ GeV) in MD;

5)$N_{h}^{MD}$ - the number of hadrons ($E_h > 30$ GeV) in MD;

6)  $N_{n}^{ND}$  - the number of thermal neutrons in the underground TND's.

Multiple parameter method is assumed to allow determination of both energy and atomic number of a
primary particle with good enough accuracy.

\section*{ACKNOWLEDGMENTS}
This is a work of the entire Carpet-3 collaboration\footnote[1]{The Carpet-3
Collaboration\\
Institute for Nuclear Research of RAS (Moscow, Russia): V.B.\,Petkov, D.D.\,Dzhappuev, Zh.Sh.\,Guliev,
A.U.\,Kudzhaev, V.I.\,Volchenko,
  G.V.\,Volchenko,  E.V.\,Gulieva, V.V.\,Alekseenko, I.A.\,Alikhanov, I.M.\,Dzaparova, A.N.\,Kurenya,
   A.F.\,Yanin, M.M.\,Kochkarov, N.F.\,Klimenko, A.S.\,Lidvansky, Yu.V.\,Stenkin, V.I.\,Stepanov, V.A.\,Kozyarivsky,
    A.B.\,Chernyaev,  R. A.\,Mukhamedshin\\
Moscow Engineering Physics Institute (Moscow, Russia): A.A.Petrukhin, N.V.Ampilogov, A.G.Bogdanov,
D.M.Gromushkin, V.V.Kindin,
                K.G.Kompaniets, S.S.Khokhlov, I.I.Yashin\\
 Kabardino-Balkarian State University (Nalchik, Russia):
A.Kh.Khokonov, M.Kh.Khokonov\\
The Andrzej Soltan Institute for Nuclear Studies (Lodz, Poland): J.Szabelski, T.Wibig, K.Jedrzejczak,
M.Kasztelan\\
South Federal University (Rostov-on-Don, Russia): Yu.S.\,Grishkan, M.A.\,Marachkov\\
University of Oulu (Oulu, Finland): T. Raiha, J. Sarkamo}.

 The work was supported by the "Neutrino physics" Program for Basic Research of the Presidium
  of the Russian Academy of Sciences. The work was supported in part by "State Program for Support of Leading
   Scientific Schools" (project no. NSh-321.2008.2) and  the Russian Foundation for Basic
Research (RFBR grants 08-02-01208 and 09-02-00293).

\end{document}